\documentclass[review]{elsarticle}

\usepackage{multirow}
\usepackage{amsmath,amssymb}
\biboptions{sort&compress}

\journal{Journal of Alloys and Compounds}
\begin{document}
\begin{frontmatter}
\title{Alloying effect on the ideal tensile strength of ferromagnetic and paramagnetic bcc iron}

\author[AMP]{Xiaoqing Li\corref{cor1}}
\ead{xiaoqli@kth.se}
\cortext[cor1]{Corresponding author}

\author[AMP]{Stephan Sch\"onecker\corref{cor1}}
\ead{stesch@kth.se}

\author[spo,klm]{Jijun Zhao\corref{cor1}}
\ead{zhaojj@dlut.edu.cn}

\author[AMP,DPA]{B\"{o}rje Johansson}
\author[AMP,DPA,RIS]{Levente Vitos}

\address[AMP]{Applied Materials Physics, Department of Materials Science and Engineering, Royal Institute of Technology, Stockholm SE-10044, Sweden}
\address[spo]{School of Physics and Optoelectronic Technology and College of Advanced Science and Technology, Dalian University of Technology, Dalian 116024, China}
\address[klm]{Key Laboratory of Materials Modification by Laser, Electron, and Ion Beams (Dalian University of Technology), Ministry of Education, Dalian 116024, China}
\address[DPA]{Department of Physics and Astronomy, Division of Materials Theory, Uppsala University, Box 516, SE-75120, Uppsala, Sweden}
\address[RIS]{Research Institute for Solid State Physics and Optics, Wigner Research Center for Physics, Budapest H-1525, P.O. Box 49, Hungary}

\begin{abstract}
Using \emph{ab initio} alloy theory formulated within the exact muffin-tin orbitals theory in combination with the coherent potential approximation, we investigate the ideal tensile strength (ITS) in the $[001]$ direction of bcc ferro-/ferrimagnetic (FFM) and paramagnetic (PM) Fe$_{1-x}M_{x}$ ($M=$ Al, V, Cr, Mn, Co, or Ni) random alloys.
The ITS of ferromagnetic (FM) Fe is calculated to be $12.6$\,GPa, in agreement with available data, while the PM phase turns out to posses a significantly lower value of $0.7\,$GPa.
Alloyed to the FM matrix, we predict that V, Cr, and Co increase the ITS of Fe, while Al and Ni decrease it. Manganese yields a weak non-monotonic alloying behavior.
In comparison to FM Fe, the alloying effect of Al and Co to PM Fe is reversed and the relative magnitude of the ITS can be altered more strongly for any of the solutes.
All considered binaries are intrinsically brittle and fail by cleavage of the $(001)$ planes under uniaxial tensile loading in both magnetic phases.
We show that the previously established ITS model based on structural energy differences proves successful in the PM Fe-alloys
but is of limited use in the case of the FFM Fe-based alloys. The different performance is attributed to the specific interplay between magnetism and volume change in response to uniaxial tension.
We establish a strong correlation between the compositional effect on the ITS and the one on the shear elastic constant $C'$ for the PM alloys and briefly discuss the relation between hardenability and the ITS.
\end{abstract}


\end{frontmatter}

\section{\label{sec:introduction}Introduction}

The mechanical properties of engineering materials are of primary interest because they determine the ability to withstand loads without failure. High strength and good ductility are characteristic of metallic materials. Both properties are controlled by the occurrence or propagation
of dislocations, cracks, grain boundaries, and other microstructural defects.
In the absence of defects, the ideal strength of a material is the strength at which a perfect crystal itself becomes unstable with respect to a homogeneous strain. This strength defines the intrinsic upper bound strength and has been accepted as an essential mechanical parameter of single crystal materials~\cite{inherent:property,study:8,Kelly:1986}.

The ideal strength is involved in the theory of fracture and the nucleation of defects~\cite{Thomson:1986,Jokl:1980,Morris:2001b}. The local inhomogeneous stress distribution close to the tip of a crack, and thus the transgranular cleavage behavior of materials, is related to the ideal tensile strength (ITS). 
The stress necessary for the nucleation of a dislocation loop can be identified with the ideal shear strength.
The ideal strength determines whether a material is intrinsically brittle or ductile~\cite{Kelly:1967,Qi:2014}. In the former case, the tensioned lattice fails by cleavage of interatomic planes in the direction perpendicular to the applied load.
Even under strict uniaxial tensile loading, an immaculate crystal may, however, not cleave but rather yield under the circumstance that a shear instability along another direction occurs first. This scenario defines intrinsically ductile materials and occurs, for example, for the bcc metals V and Nb loaded along the $[001]$ direction, where a shear instability is activated in the \{112\}$\langle 111\rangle$ slip system before the cleavage strength of the $(001)$ planes is reached~\cite{study:8,Cerny:2013}.

Further, the ideal strength is relevant in situations where dislocation activity is suppressed or for systems with very low defect density. The plastic deformation behavior of a multicomponent solid solution based on the Ti-Nb binary (``gum metals'') is believed to be governed by the ideal strength rather than conventional dislocation mobility~\cite{Li:2007}.
The second mentioned situation is frequently met in the deformation behaviors of filamentary crystals or graphene sheets, the mechanical properties of nanowires and nanopillars, and in nanoindentation experiments on thin films and nanostructures~\cite{Kelly:1986,Brenner:1956,Lowrya:2010,Lee:2008,Ju:2010}.
In some cases the measured maximum strength approximates the theoretically predicted ideal strength. For example, the maximum shear strength of Mo has been measured by nanoindentation on single-crystalline nanopillars~\cite{Lowrya:2010}. The experimental value is $15.8-16.7$\,GPa, which is close to the theoretical prediction $17.6-18.8$\,GPa~\cite{Krenn:2002}.

In recent years, considerable attention has been paid to the theoretical determination of the large-strain response and the ITS of elemental solids and intermetallic compounds by \emph{ab initio} methods~\cite{study:2,study:4,study:5,study:6,study:7,study:8}.

Among the bcc metals, elemental Fe as the basic ingredient in steel received great scientific interest~\cite{inherent:property,Cerny:2007,Liu,idealandmagnetic:2002,Sob:2004,Cerny:2013,Li:2014,Li:2015b}.
The picture that has evolved for FM ordered Fe is that the $[001]$ direction is the weakest one in response to uniaxial stress. The deformation is governed by a primary tetragonal deformation path, but a bifurcation to a secondary orthorhombic deformation path, triggered by a shear instability, occurs after the ITS on the primary path is reached. Accordingly, bcc Fe strained along the $[001]$ direction fails by cleavage as opposed to V and Nb~\cite{study:8,Cerny:2013}.
If the failure mode in tension were restricted to cleavage of the $(001)$ plane, the ITS of Fe turned out to be the lowest among all bcc transition metals~\cite{Mo:1,study:8,inherent:property,Cerny:2007}.

Alloying plays an essential role in material and steel design as the alloy properties are usually superior to those of the pure constituents. As regards the ideal strength, tuning its magnitude and the ductile versus brittle failure mode are of obvious interest.
The \emph{ab initio} description of the ITS in substitutionally disordered alloys is, however, rather limited. One early attempt~\cite{Li:2007} employed the virtual-crystal approximation (VCA) to study the ITS of the Ti-V binary.
The applicability of the VCA is by construction restricted to binary solid-solutions composed of neighboring elements in the periodic table, which covers a tiny fraction of the physically interesting cases. A more general approach to multi-component random solid solutions is given by the coherent-potential approximation (CPA)~\cite{cpa:1}.
Recently, Li \emph{et al.}~\cite{Xiaoqing:2013} investigated the composition-dependent ITS of bcc V-Cr-Ti solid solutions demonstrating the applicability of the CPA for this type of calculations. The V-Cr-Ti ternary is presently considered as one of the primary structural materials for the first wall and blanket structure of future fusion reactor~\cite{candidate:1,candidate:2}.

In a recent study~\cite{Li:2014}, we focused on the anomalously low ITS of Fe compared to other bcc transition metals, which can partially be ascribed to its weak FM behavior. By means of analyzing electronic structure and magnetism, we found that a small amount of alloying addition can turn the weak FM behavior in the Fe host more stable against structural deformations, resulting in anomalous alloying trends.
The purpose of this paper is to give an in-depth investigation of the ITS of bcc Fe-based alloys in magnetically ordered [ferro-/ferrimagnetic (FFM)] and disordered [paramagnetic (PM)] phases and to establish a relation to other mechanical properties.
Our results give useful limits on the attainable combination of strength and toughness of Fe-based alloys and provide a consistent theoretical guide to further optimization of the composition of Fe-based alloys in multiscale-materials design~\cite{Guo:2001,Morris:2001b}.

\section{\label{sec:computationalmethod}Computational method}

\subsection{Total energy calculations}

The \emph{ab initio} approach used in this work is based on density-functional theory (DFT)~\cite{DFT}. We adopted the generalized-gradient approximation (GGA) of the Perdew-Burke-Ernzerhof (PBE)~\cite{PBE} functional to describe exchange and correlation, which is well known to give the correct FM bcc ground state for Fe.
For the main part of the work, the Kohn-Sham equations were solved using the exact muffin-tin orbitals (EMTO) method~\cite{EMTO:1,EMTO:3}. The problem of chemical disorder was treated within the CPA and the total energy was computed via the full charge-density technique~\cite{cpa:1,cpa:3,cpa:4}.
We point out that, due to its single-site nature, the CPA describes a homogeneous random solid solution and omits the inter-dependence between magnetic state and local environment. The magnetic state of the presently considered solute atoms in the FM Fe host is determined by the coherent Green function rather than by actual neighboring atoms.

The EMTO method is an improved screened Korringa-Kohn-Rostoker method~\cite{EMTO:1}, where the full potential is represented by overlapping muffin-tin potential spheres. The potential is spherically symmetric inside these spheres and constant in between. By using overlapping spheres, one describes more accurately the exact crystal potential compared to conventional muffin-tin methods. 
The accuracy of the EMTO method for the equation of state and elastic properties of metals and alloys was demonstrated in a number of previous works~\cite{Xiaoqing:2013,work:7,Xiao:2012} and is assessed for the ITS of Fe in Sec.~\ref{sec:equilibrium}.

The PM state well above the magnetic transition temperature (where the magnetic short range order effects are negligible) was described by the disordered-local moment (DLM) model~\cite{Gyorffy:1985} in the CPA framework. Within the DLM picture, PM Fe and Fe$_{1-x}M_{x}$ binary alloys were simulated as a binary alloy Fe$\uparrow$Fe$\downarrow$ and a quaternary (Fe$\uparrow$Fe$\downarrow$)$_{1-x}$(\emph{M}$\uparrow$\emph{M}$\downarrow$)$_{x}$ alloy, with an equal amount of spin-up ($\uparrow$) and spin-down ($\downarrow$) alloy components, respectively.

Additional test calculations for FM Fe were performed with the full-potential local orbital scheme FPLO-9~\cite{Koepernik:1999}, the projector-augmented wave (PAW) method as implemented in VASP release 5.3.3~\cite{Kresse:1996}, and the full-potential linearized augmented-plane wave (FP-LAPW) code ELK version 2.2.9~\cite{Elk:online} using PBE.
The FPLO calculations were performed in the scalar-relativistic approximation. Convergence of numerical parameters, in particular integration meshes and the basis, was carefully checked. Linear-tetrahedron integrations with Bl\"ochl corrections were done on a $24\times 24\times 24$ mesh in the full Brillouin zone and the valence basis comprised $3d$, $4spd$, and $5s$ states augmented by $5pd$ and $6s$ polarization states.
The ELK scalar-relativistic calculations employed the default valence basis set. The muffin-tin radius was fixed to 2.100\,Bohr for all calculations ensuring that there is no overlap between neighboring spheres. Brillouin zone integrations were done on a $18\times18\times18$ $k$-point mesh smeared by a Fermi-Dirac function and a small smearing parameter $5\times 10^{-4}$\,Ha.
The augmented plane-wave cutoff was set to 9.2\,Ha, the angular momentum cut-off for the APW functions was set to 10, and the angular momentum cut-off the outer-most loop in the Hamiltonian and overlap matrix setup were set to 8 to ensure converged results.
All VASP calculations were done with the default $spd$-valence state PAW potential and the global 'Accurate' precision switch. Total energies and the computed stress tensor were found converged for a plane-wave cutoff $\ge 500$\,eV. The unit cell was relaxed until residual stresses perpendicular to the applied force were smaller than $1\times 10^{-3}$\,eV/\AA.
Brillouin zone integrations were done on a $24\times24\times24$ Monkhorst-Pack mesh, smeared by a first order Methfessel-Paxton scheme with smearing parameter 0.2\,eV. The grid for augmentation charges contained eight times more points than default.

\subsection{Ideal tensile strength calculation for bcc crystals}

The principles of the response of bcc crystals to uniaxial loading were developed in a series of works by Milstein \emph{et al.}~\cite{Milstein:1970,Milstein:1971,Milstein:1988}.
Since $[001]$ was already identified as the weakest direction of bcc Fe~\cite{Cerny:2007,Sob:2004,Cerny:2013,Liu}, here we concentrate on this direction and compute the ITS of bcc Fe and Fe-based alloys upon uniaxial loading along the [001] direction. Assuming an uniaxial tensile load, the tensile stress $\sigma (\varepsilon)$ can be calculated by~\cite{stress}
\begin{equation}
\sigma(\varepsilon)=\frac{1+\varepsilon}{\Omega(\varepsilon)}\frac{\partial (\Delta E)}{\partial\epsilon}.
\label{eq:stress}
\end{equation}
$\Delta E$ and $\Omega(\varepsilon)$ are the uniaxial strain energy and the volume at a given tensile strain ($\epsilon$).
The first maximum on the stress-strain curve defines the ITS ($\sigma_{\textrm{m}}$) with corresponding maximum strain ($\epsilon_\textrm{m}$).

Uniaxial loading along the $[001]$ direction reduces the symmetry of the bcc lattice to body-centered tetragonal (bct, with lattice parameter of quadratic basal plane and height of cell denoted by $a$ and $c$, respectively) on the primary deformation path (Bain path). Clatterbuck \emph{et al.}~\cite{inherent:property} reported a bifurcation from the primary deformation path to a secondary orthorhombic deformation path in FM Fe. The branching occurs, however, for strains well above $\epsilon_\textrm{m}$. Thus, an isotropic Poisson contraction along the Bain path describes appropriately the symmetry of the distorted Fe lattice up to $\epsilon_\textrm{m}$.
The branching point may, however, shift towards strains smaller than the maximum strain along the primary deformation path as a result of alloying or due to different magnetic order in Fe (PM Fe host). We account for this possibility in the present work.

On the primary tetragonal deformation path, the strain energy $\Delta\emph{E}(c;[001])$ is described by the total energy change upon elongating the bcc crystal in the $[001]$ direction, and relaxing it with respect to the dimensions in the $(001)$ plane, \emph{viz.}
\begin{equation}
\Delta\emph{E}(c;[001])= \min_{a} E(a,c) - E_0,
\label{eq:energy:001}
\end{equation}
where the initial, undistorted state corresponds to the equilibrium bcc structure with energy $E_0$ and $c_0=a_0$, and $\epsilon$ is related to $c$ by $c=c_0(1+\epsilon)$. At $c/a=\sqrt{2}$, the bct lattice coincides with the face-centered cubic (fcc) lattice.
On the secondary orthorhombic (orth) deformation path, we consider
\begin{equation}
\Delta\emph{E}(c_{\textrm{orth}};[001]) = \min_{a_{\textrm{orth}},b_{\textrm{orth}}} E(a_{\textrm{orth}},b_{\textrm{orth}},c_{\textrm{orth}}) - E_0,
\end{equation}
minimizing the total energy with respect to two lattice parameters of the face-centered orthorhombic lattice, $a_{\textrm{orth}}$ and $b_{\textrm{orth}}$. Notice that to describe the bifurcation from the primary tetragonal to the secondary orthorhombic deformation path, the face-centered tetragonal reference frame of the Bain transformation may be used instead of the bct reference frame. For a more detailed technical description of ideal tensile strength simulations with the EMTO method, we also refer the reader to Ref.~\cite{Xiaoqing:2013}.

The ideal tensile strength is obtained from the maximum of the $\sigma(\epsilon)$ curve. In practice, we fit both the total energy $\Delta\emph{E}(\epsilon)$ and the volume $\Omega(\epsilon)$ by a low-order polynomial as a function of $\epsilon$.  Based on a careful analysis, we found that the \emph{relative} errors originating from the numerical fitting is below $\sim 2\%$ for $\sigma_{\textrm{m}}$ and $\sim 3\%$ for $\epsilon_\textrm{m}$. The somewhat larger error in the ideal strain is due to the fact that around the inflection point of $\Delta E$ the total energy behaves almost linearly as a function of $\epsilon$. Hence the slope of $\Delta E(\epsilon)$ is rather robust  but the actual position of the inflection point is more sensitive to the details of the fit function.

\section{\label{sec:resultsanddiscussion}Results}
\subsection{\label{sec:equilibrium}Ideal tensile strength of bcc iron}
To assess our computational approach, we first performed the simulation of a tensile test in bcc Fe for uniaxial loading along the $[001]$ direction. The present theoretical Wigner-Seitz radius ($w$) of bcc FM Fe is $2.637\,\textrm{Bohr}$, which agrees well with the results from FPLO ($2.633\,\textrm{Bohr}$), ELK ($2.638\,\textrm{Bohr}$), VASP ($2.636\,\textrm{Bohr}$),  and underestimates the extrapolated, low-temperature experimental Wigner-Seitz radius ($2.668\,\textrm{Bohr}$~\cite{Villars:1991}) slightly by $1.2\,\%$.

In equilibrium, the stable magnetic order in bcc Fe is FM. The magnetic order of the strained Fe lattice may, however, change along the uniaxial loading process, which is for strains up to the branching point governed by the Bain transformation as discussed in the previous section. The fcc state of Fe lying on the Bain transformation path (albeit at strains much larger than $\epsilon_{\textrm{m}}$) exhibits a non-collinear spin arrangement as measured in fcc Fe precipitates and in thin fcc Fe films~\cite{Tsunoda:2007,Meyerheim:2009}. That indicates that the ground state magnetic order of the strained bct Fe lattice begins to differ from the FM order at some particular strain in the range between bcc and fcc along the primary transformation path. If this would be the case for $\epsilon \le \epsilon_{\textrm{m}}$, then additional magnetic orders should be considered.
However, there are strong indications that the FM order is the prevailing magnetic state for strains smaller than and somewhat above $\epsilon_{\textrm{m}}$~\cite{idealandmagnetic:2002,Tsetseris:2005,Friak:2001}.

Clatterbuck \emph{et al.}~\cite{idealandmagnetic:2002} computed the ITS of Fe in the $[001]$ direction considering FM order, a collinear anti-ferromagnetic structure (AFM, magnetic moment sequence $\uparrow \downarrow$), and a collinear double layer anti-ferromagnetic structure (DAFM, magnetic moment sequence $\uparrow \uparrow \downarrow\downarrow$). Their results show that Fe remains FM up to the point of its elastic instability during uniaxial tension, which lies at approximately $\epsilon_\textrm{m}=15\,\%$ with $c/a\approx 1.16$ and $\Omega/\Omega_{\textrm{exp}}> 1$ ($\Omega_{\textrm{exp}}$ denotes the experimental atomic volume of bcc Fe).
Further evidence is given by Tsetseris~\cite{Tsetseris:2005} and Fri\'{a}k \emph{et al.}~\cite{Friak:2001} who published minimum energy contour plots with respect to various magnetic orders (FM, AFM, and DAFM order by Fri\'{a}k \emph{et al.} and non-collinear magnetism via a spin spiral formalism by Tsetseris) as a function of the bct geometry, thereby defining magnetic phase boundaries between different magnetic states. According to these references, the FM state is the predominant magnetic order in the configuration space for $c/a\le 1.25$ and $\Omega/\Omega_{\textrm{exp}}\ge 0.95$. Both references hence indicate that the point of elastic instability of Fe reported by Clatterbuck \emph{et al.} is indeed located far from the borderline of FM order towards any other investigated magnetic state.

Hence, here we assume that the FM ordered state of the bcc Fe matrix remains FM ordered during the entire deformation process.

\begin{table}[thb]
\centering
\caption{\label{table:one}The ideal tensile strength $\sigma_{\textrm{m}}$ and the corresponding strain $\epsilon_{\textrm{m}}$ in the $[001]$ direction of FM and PM bcc Fe. The present results (EMTO, FPLO, PAW and FP-LAPW) are compared with previous PAW~\cite{inherent:property,Cerny:2007,Liu} and FP-LAPW~\cite{Sob:2004} data for FM Fe. All quoted references employed GGA functionals.}
\begin{tabular}{llcc}
\hline
  \multirow{2}{*}{state}& \multirow{2}{*}{method} & \multicolumn{2}{c}{direction/$[001]$}\\
\cline{3-4}
 &    & $\sigma_{\textrm{m}}$ (GPa) & $\epsilon_{\textrm{m}} (\%)$ \\
 \hline
 \multirow{8}{*}{FM Fe} &   EMTO (this work) & 12.6 & 14.1 \\
&     FPLO (this work)& 13.0 & 15.2 \\
&     FP-LAPW (this work)& 12.8 & 14.3 \\
&     PAW (this work)& 11.6 & 14.7 \\
&     PAW Ref.~\cite{inherent:property} & 12.6 & 15 \\
&     PAW Ref.~\cite{Cerny:2007} & 12.4 & 16  \\
&     PAW Ref.~\cite{Liu}  & 12.4 & 14  \\
&     FP-LAPW Ref.~\cite{Sob:2004}  & 12.7 & 15  \\
     \hline
 PM Fe & EMTO (this work) &0.7&2.8\\
      \hline
\end{tabular}
\end{table}
The ITS $\sigma_{\textrm{m}}$ for the FM and PM states of bcc Fe and corresponding strain $\epsilon_{\textrm{m}}$ from this work and available literature are compared in Table~\ref{table:one}. For FM Fe, the present results are consistent with the available literature results.
Furthermore, the sum of these data allows to establish a scatter typical for this type of calculation.
Three previous PAW works using VASP gave similar values for $\sigma_{\textrm{m}}$~\cite{inherent:property,Cerny:2007,Liu}, but scatter by $\pm1\,\%$ around the mean value of $\epsilon_{\textrm{m}}$ ($15\,\%$). The presently computed ITS using VASP is somewhat lower than these former PAW results. The likely reason is that all previous PAW data were obtained with a lower plane-wave energy cut-off (smaller than $350$\,eV) than the converged value used in this work ($500$\,eV). Decreasing the cut-off to $350$\,eV, we found that the ITS of Fe  increases to $12.1$\,GPa indeed bringing it in closer agreement with these previous assessments.

Comparing the two magnetic phases, we infer that PM Fe possesses significantly lower $\sigma_{\textrm{m}}$ and $\epsilon_{\textrm{m}}$. This fact is a further manifestation of the impact of the magnetic state on the mechanical strength of bcc Fe, which is, for example, also observed (albeit to a lesser extent) for the elastic constants~\cite{Adams:2006}. Here we must emphasize that a proper description of the PM state is indispensable for the above results. Previous \emph{ab initio} calculations performed for non-magnetic Fe lead to a mechanically unstable bcc phase~\cite{Herper:1998} and thus the ITS could not be defined. It is the disordered magnetic state with non-vanishing local magnetic moments which stabilizes the bcc PM Fe mechanically~\cite{Gyorffy:1985,Li:2015b}.

Turning to the analysis of the failure mode for both magnetic states, we found that both FM and PM Fe fail by cleavage under $[001]$ tension.
We obtained $c/a\approx 1.18$ and $\Omega/\Omega_{\textrm{exp}}\approx 1.04$ for the failure point of FM Fe in agreement with the aforementioned values from Clatterbuck \emph{et al.}~\cite{idealandmagnetic:2002}.
Finally, we computed that the bct to orth branching in FM Fe occurs at $\epsilon_{\textrm{orth}}=17\,\%$, i.e., well above $\epsilon_{\textrm{m}}=14.1\,\%$. This is in accordance with Ref.~\cite{inherent:property}, where the branching was reported to occur at $18$\,\% strain.

Based on the above comparison, we conclude that EMTO is able to describe the ITS of bcc Fe with an accuracy comparable with previous and present full-potential calculations.

\subsection{\label{sec:strengthFe}Ideal tensile strength of bcc Fe-based alloys}

In the following we turn to bcc Fe-based alloys and investigate the effect of alloying on the ITS of Fe$_{1-x}M_{x}$ random solid solutions, where $M=$ Al, V, Cr, Mn, Co, or Ni.
The selected solute atoms are common in commercial Fe-based steel alloys and represent a simple metal (Al), nonmagnetic (V) and magnetic (Cr, Mn, Co, and Ni) transition metals. The concentration of the solutes was varied from 0 to 10\,\% except for Mn alloyed to FM Fe where the maximum concentration was 5\,\% due to limited solubility.

In their equilibrium crystal structures, Cr and Mn exhibit (complex) AFM ground state orders while Co and Ni order FM. In the present work, we only account for collinear magnetic states, \emph{i.e.}, the magnetic moment of the solutes can be oriented parallel or antiparallel to the one of the FM Fe host giving rise to FM or ferrimagnetic order, respectively.
For the present Fe-based alloys in the bcc ground state and all considered concentrations, we found that the magnetic moments of Cr and Mn are oriented against those of Fe, while Co and Ni moments align parallel to the Fe moment. In Fe$_{1-x}$Al$_{x}$ and Fe$_{1-x}$V$_{x}$, Fe induces a small magnetic moment in Al (0.1-0.2\,$\mu_{\textrm{B}}$) and a somewhat larger moment in V ($\sim 1.0\,\mu_{\textrm{B}}$) both oriented antiparallel with respect to the magnetic moment of Fe. The same relative orientation between Fe and the solutes was found to prevail in the tensioned structures.
For the sake of simplifying the notation we refer to the set of here considered binary alloys with a FM Fe matrix as FFM Fe-alloys (for ferro-/ferrimagnetic) in the following.

\begin{figure}[tbh]
\begin{center}
\begin{tabular}{@{}ccc@{}}
\resizebox{0.5\columnwidth}{!}{\includegraphics[clip]{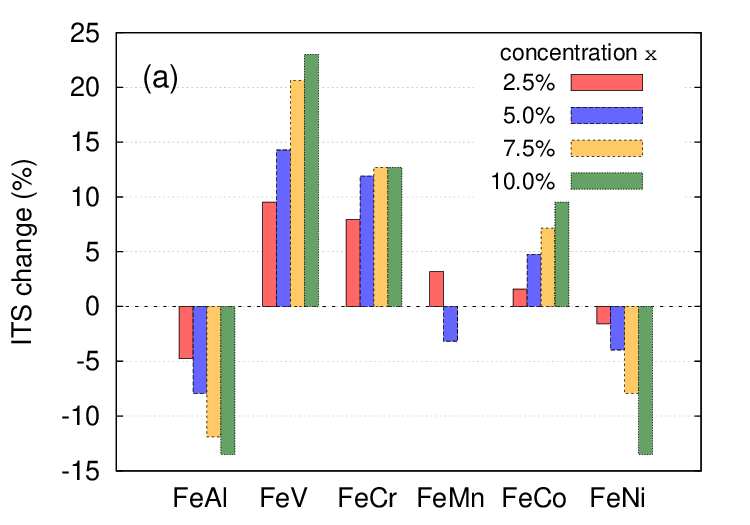}}&
\resizebox{0.5\columnwidth}{!}{\includegraphics[clip]{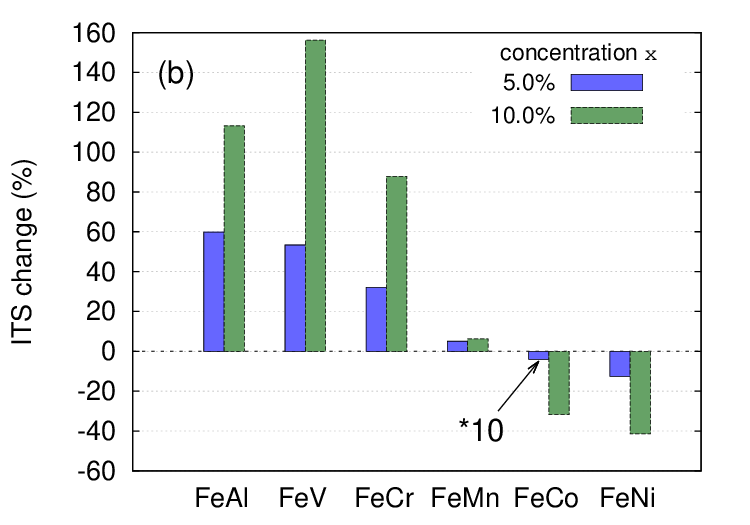}}\\
\end{tabular}
\caption{\label{fig:alloysstress}The relative ITS change with respect to the ITS of FM and PM Fe for the (a) FFM and (b) PM Fe$_{1-x}M_{x}$ alloys, respectively. The value for PM Fe$_{0.95}$Co$_{0.05}$ was multiplied by ten to ease the comparison.}
\end{center}
\end{figure}

Figure~\ref{fig:alloysstress} shows the relative change of the ITS for the six binary alloys with respect to the one of bcc Fe in FM and PM phases. The corresponding numerical data for selected compositions are listed in Tables~\ref{table:alloys stress} and~\ref{table:PMalloys stress}.
From Fig.~\ref{fig:alloysstress}(a), we can see that the calculated ITS increases with Cr, Co, or V and decreases with Ni or Al addition to FM Fe. For instance, when $10\,\%$ Cr, Co, or V is added to bcc FM Fe, the ITS of Fe increases by $12.7\,\%$, $9.5\,\%$ and $23.0\,\%$, respectively. If however $10$\,\% Ni or Al is added to the FM Fe matrix, the ITS reduces by $13.5\,\%$.
The ITS of Fe$_{1-x}$Mn$_{x}$ increases by $3.2\,\%$ as the Mn concentration increases from $0$ to $2.5\,\%$, and then reduces by $6.2\,\%$ when $5\,\%$ Mn is added to FM Fe.
In comparison to the FM state, we found that the alloying effect of Al and Co to PM Fe is reversed, \emph{i.e.}, the addition of Al (Co) increases (decreases) the ITS of PM Fe. Manganese added to PM Fe enhances the ITS only slightly.
According to Fig.~\ref{fig:alloysstress}, one infers that V is overall the most efficient alloying agent producing a biggest enhancements of the ITS for all investigated concentrations in both magnetic states.

Three important observations may be pointed out on the basis of the computed ITS data. First, fractional alloying effects on the ITS of Fe in the PM state exceed those in the FM state. In other words, the relative magnitude of the ITS can be tuned more sensitively (using lower solute concentrations) in the PM state. Second, certain solutes (Al, Co, and  5\,\% Mn) increase the ITS of Fe in one magnetic state while decreasing it in the other. Consequently, other alloy additions (V, Cr, and Ni) modify the ITS in the same manner independent on the prevalent magnetic state. Third, eye-catching among the transition metal solutes is the fact that elements with $d$-occupation number larger (smaller) than Fe decrease (increase) the ITS of PM Fe, while no similar trend appears for the alloys in the FFM phase.
Interestingly, Zhang \emph{et al.} reported the same alloying behavior (upward versus downward trend) for the solutes Al, V, Cr, Mn, Co, or Ni on the shear elastic constant $C'$ of PM bcc Fe~\cite{Zhang:2012PM}, \emph{i.e.}, Co and Ni lower $C'$ of Fe while the other elements stiffen $C'$.
A possible correlation between the ITS and $C'$ is investigated in Sec.~\ref{sec:correlation}.

\begin{table*}[tbh]
\centering
\caption{\label{table:alloys stress}Ideal tensile strength ($\sigma_{\textrm{m}}$, in GPa) calculated in the $[001]$ direction for FFM Fe$_{1-x}M_{x}$ alloys. For pure Fe, $\sigma_{\textrm{m}} = 12.6$\,GPa.}
\begin{tabular}{l*{5}{c}lc}
\hline
\multirow{2}{*}{$x$}  & & &$\sigma_{\textrm{m}}(x)$ &&& \multirow{2}{*}{$x$} & $\sigma_{\textrm{m}}(x)$ \\
\cline{2-6}\cline{8-8}
 & Fe-Al & Fe-V & Fe-Cr& Fe-Co & Fe-Ni & & Fe-Mn \\
 \hline
 0.025 &12.0&13.8&13.6&12.8&12.4& 0.0125& 12.8     \\
 0.05  &11.6&14.4&14.1&13.2&12.1& 0.025 & 13.0\\
 0.075 &11.1&15.2&14.2&13.5&11.6& 0.0375& 12.6  \\
 0.1   &10.9&15.5&14.2&13.8&10.9& 0.05&  12.2  \\
 \hline
\end{tabular}
\end{table*}

\begin{table*}[tbh]
\centering
\caption{\label{table:PMalloys stress}Ideal tensile strength ($\sigma_{\textrm{m}}$, in GPa) calculated in the $[001]$ direction for PM Fe$_{1-x}M_{x}$ alloys. For pure Fe, $\sigma_{\textrm{m}}=0.7$\,GPa.}
\begin{tabular}{l*{6}{c}}
\hline
\multirow{2}{*}{$x$}  &&&& $\sigma_{\textrm{m}}(x)$\\
\cline{2-7}
 & Fe-Al & Fe-V & Fe-Cr& Fe-Mn& Fe-Co & Fe-Ni\\
 \hline
 0.05  &1.12&1.08&0.93&0.74&0.68&0.61\\
 0.1   &1.50&1.80&1.32&0.75&0.48&0.41\\
 \hline
\end{tabular}
\end{table*}

Focusing on the analysis of the failure mode for the present Fe$_{1-x}M_{x}$ alloys, we found that for all binaries and concentrations considered here branching from the primary bct to the secondary orthorhombic deformation path occurs at strains larger than $\epsilon_{\textrm{m}}(x)$ corresponding to the ITS of alloy Fe$_{1-x}M_x$.
In other words, all binaries within the considered compositional interval are intrinsically brittle and fail by cleavage of the $(001)$ planes under uniaxial tensile loading along $[001]$. 

We should mention that the chemical effect of Co on the ITS of bcc Fe for both magnetic phases is rather modest. This finding might look contradictory when compared to the Co effect on the thermal stress of bcc Fe~\cite{Li:2015b}. The reason is that here the calculations focus merely on the intrinsic alloying effects and do not consider the gradual loss of the magnetic order with increasing temperature. The previously predicted and discussed huge Co effect on the ITS between $\sim 600$-1000\,K is due to the fact that Co increases the Curie temperature and thus brings the Fe-Co alloy back into the FM 'regime'.

\section{\label{sec:discussions}Discussion}

\subsection{\label{sec:discussions:sed}The structural energy difference model for the ITS}

In this section, we make an attempt to describe and predict the alloying effect on the ITS of Fe-based alloys using accessible physical quantities. To this end, we start from a previously established model applicable to bcc systems which is based on structural energy differences (SEDs):
originally proposed for binary and ternary vanadium based V$_{1-y-z}$Cr$_{y}$Ti$_{z}$ random solid solutions ($0\le y+z\le 0.1$), we correlated the alloying induced change of the ITS on the primary tetragonal deformation path with that of the nearest minimum-energy barrier (saddle point) on a \emph{constrained} strain-energy curve~\cite{Xiaoqing:2013}.
Specifically, the maximum stress ($\sigma^{\phantom{S}}_{\textrm{m}}$) along the $[001]$ direction was correlated with $\sigma^{\textrm{SED}}_{\textrm{m}}$ defined as
\begin{equation}
\sigma^{\textrm{SED}}_{\textrm{m}} = \frac{1+\Delta\epsilon}{\Omega_{\textrm{bcc}}}\frac{\Delta E^{\textrm{SED}}}{\Delta \epsilon}.
\label{eq:stress:strdiff}
\end{equation}
Here, $\Delta E^{\textrm{SED}}\equiv E_{\textrm{saddle}} - E_{\textrm{bcc}}$ denotes the energy difference between saddle point and ground state (that is the height of the barrier) referred to as SED. The SED is evaluated at fixed bcc ground state volume $\Omega_{\textrm{bcc}}$ and $\Delta \epsilon$ is the strain necessary to transform the bcc lattice into the bct configuration of the saddle point along the constant volume Bain transformation path. In the case of V$_{1-y-z}$Cr$_{y}$Ti$_{z}$, the saddle point arises due to the fcc structure on the strain-energy curve (at $\Delta \epsilon\approx 0.260$) and $\Delta E^{\textrm{SED}}$ coincides with $E_{\textrm{fcc}} - E_{\textrm{bcc}}$~\cite{Xiaoqing:2013}.
The physical motivation for model~\eqref{eq:stress:strdiff} is based on the fact that the uniaxial strain energy must level off at the saddle point, which implies a limitation on the maximum stress since it restricts $\Delta\emph{E}(c;[001])$ to approximately the SED within the strain interval to accomplish the transformation from bcc to the saddle point configuration, i.e., the ratio $\Delta E^{\textrm{SED}}/\Delta \epsilon$ is bounded.

The alloying trend is captured by Eq.~\eqref{eq:stress:strdiff} if there is a correlation between the change of the ITS as a function of concentration, $\Delta\sigma_{\textrm{m}}(x) \equiv \sigma_{\textrm{m}}(x)-\sigma_{\textrm{m}}(x_0)$, and the change of the SED, $\Delta(\Delta E^{\textrm{SED}})(x) \equiv \Delta E^{\textrm{SED}}(x)-\Delta E^{\textrm{SED}}(x_0)$, where $x_0$ is the reference concentration.
We show in the following that the correlation suggested by Eq.~\eqref{eq:stress:strdiff} has limited applicability for the presently considered FFM Fe-based alloys as opposed to the PM Fe-based alloys and the previously studied V$_{1-y-z}$Cr$_{y}$Ti$_{z}$ ternary~\cite{Xiaoqing:2013}.

Previous investigations for pure Fe revealed that the fcc structure is the nearest saddle point configuration to the bcc phase on the uniaxial strain energy curve if the magnetic state is constrained to FM order~\cite{inherent:property,Liu,idealandmagnetic:2002,Sob:2004}.
We found that the saddle point state also coincides with the fcc structure for the presently investigated FFM Fe-based alloys. For the PM Fe and Fe-based alloys, the saddle point configuration has bct symmetry with axial ratio $1< c/a < \sqrt{2}$.
Hence, the constant volume SED is given by $\Delta E^{\textrm{SED}} = E_{\textrm{bct}} - E_{\textrm{bcc}}$ for all investigated PM systems, while $\Delta E^{\textrm{SED}} = E_{\textrm{fcc}} - E_{\textrm{bcc}}$ for the FFM binaries.

\begin{figure}[tbh]
\begin{center}
\begin{tabular}{@{}ccc@{}}
\resizebox{0.5\columnwidth}{!}{\includegraphics[clip]{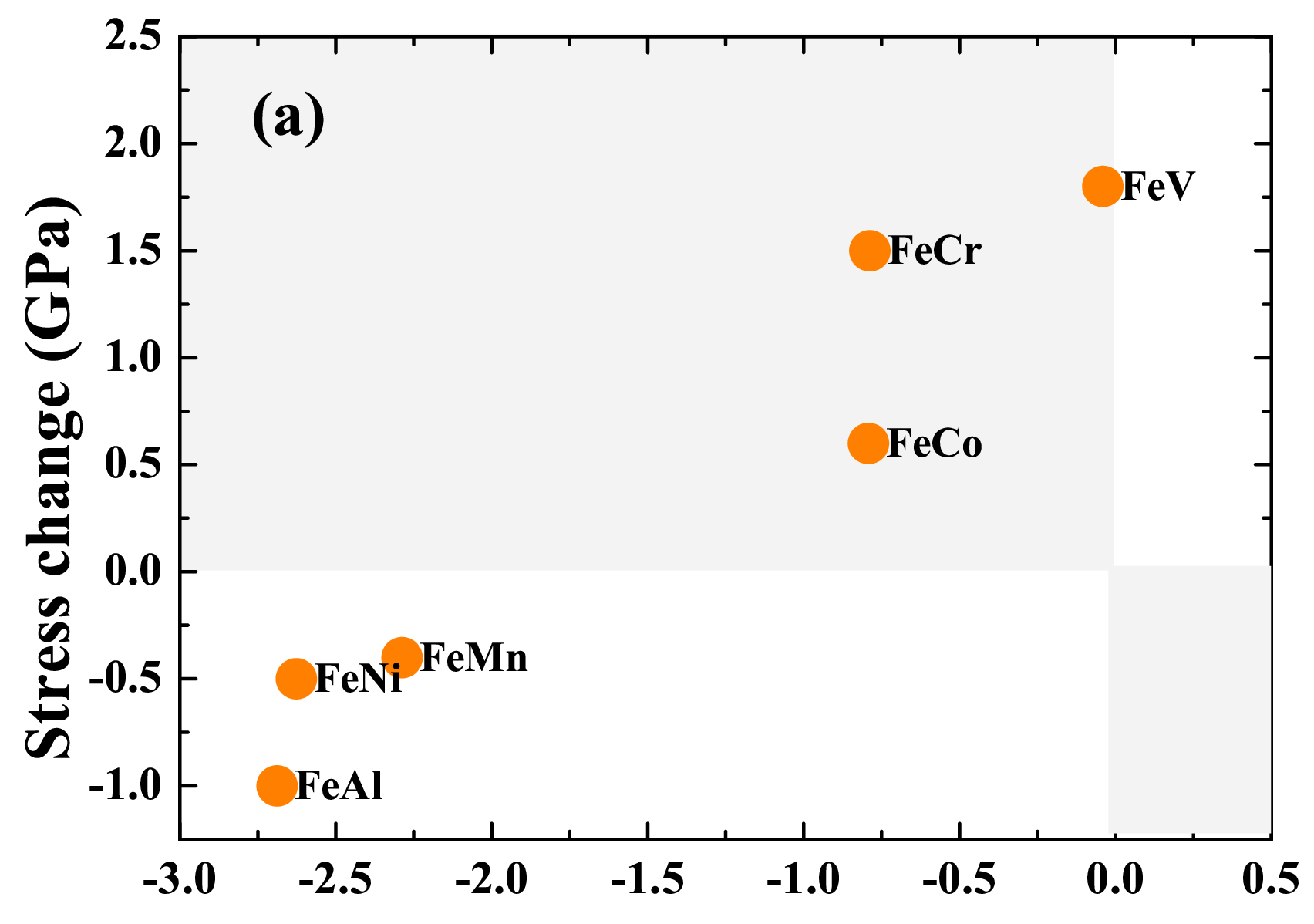}}\\
\resizebox{0.5\columnwidth}{!}{\includegraphics[clip]{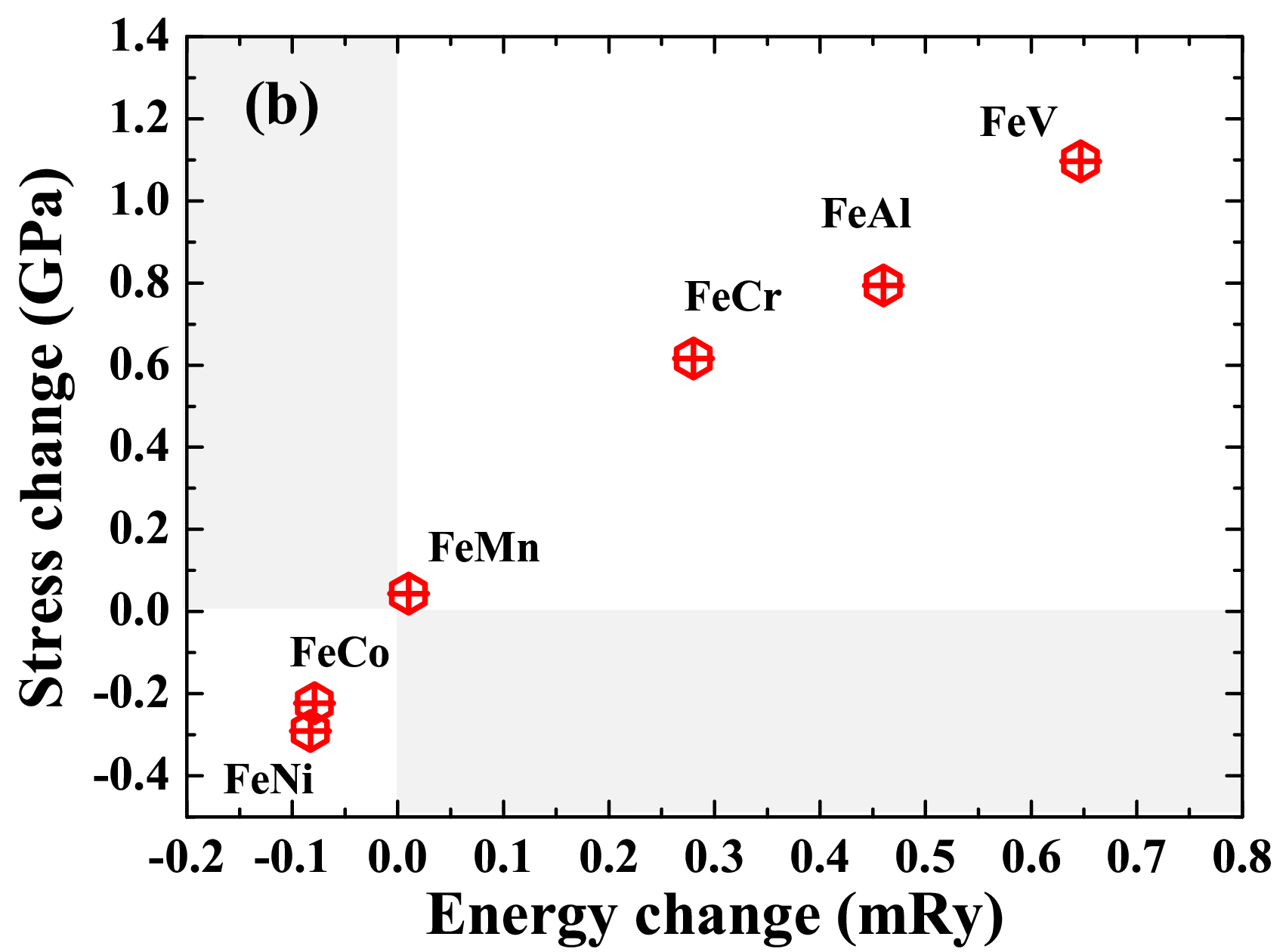}}\\
\end{tabular}
\caption{\label{fig:FM}The change of the ITS, $\Delta \sigma_{\textrm{m}}$, versus the change of the fcc-bcc SED and bct-bcc SED $\Delta(\Delta E^{\textrm{SED}})$, for Fe$_{1-x}M_{x}$ alloys in (a) the FFM state for a concentration increase from 0\,\% to 5\,\% and in (b) the PM state for a concentration increase from 0\,\% to 10\,\%, respectively. Data in unshaded areas affirm a correlation based on Eq.~\eqref{eq:stress:strdiff}. }
\end{center}
\end{figure}

Figure~\ref{fig:FM} displays $\Delta\sigma_{\textrm{m}}(x)$, obtained from the values in Tables~\ref{table:alloys stress} and~\ref{table:PMalloys stress}, as a function of $\Delta(\Delta E^{\textrm{SED}})(x)$ calculated with EMTO-CPA for both magnetic states. The alloying effects were obtained by increasing the concentration of the solute from 0\,\% to 5\,\% and from 0\,\% to 10\,\% (\emph{i.e.}, $x_{0}=0$ and $x=0.05/0.1$) for FFM and PM alloys, respectively.
For the FM state [Fig.~\ref{fig:FM}(a)], we can see that Ni, Mn, and Al decrease both the ITS and the SED, while Co, Cr, and V increase the ITS but decrease the SED.
We also investigated the correlation between $\sigma_{\textrm{m}}(x)$ and $\Delta(\Delta E^{\textrm{SED}})(x)$ increasing the concentration of the solute from $x_0= 0.05$ to $x=0.1$, but the result is qualitatively identical to the one depicted in Fig.~\ref{fig:FM}(a).
From Fig.~\ref{fig:FM}(b), one can see that $\Delta\sigma_{\textrm{m}}(x)$ correlates strongly with the $\Delta(\Delta E^{\textrm{SED}})(x)$ in the PM state.
From these results, we infer that the correlation between the SEDs and the ITSs suggested by Eq.~\eqref{eq:stress:strdiff} is not supported for the FFM Fe-based alloys, while it holds for the PM Fe-based alloys.

In the following section, we show that the above failure can be traced back to the effect of magnetism.

\subsection{\label{sec:magneffects}Magnetic effects on the ITS}

\begin{figure*}[hbtc]
\begin{center}
\begin{tabular}{@{}cc@{}}
\resizebox{!}{4cm}{\includegraphics[clip]{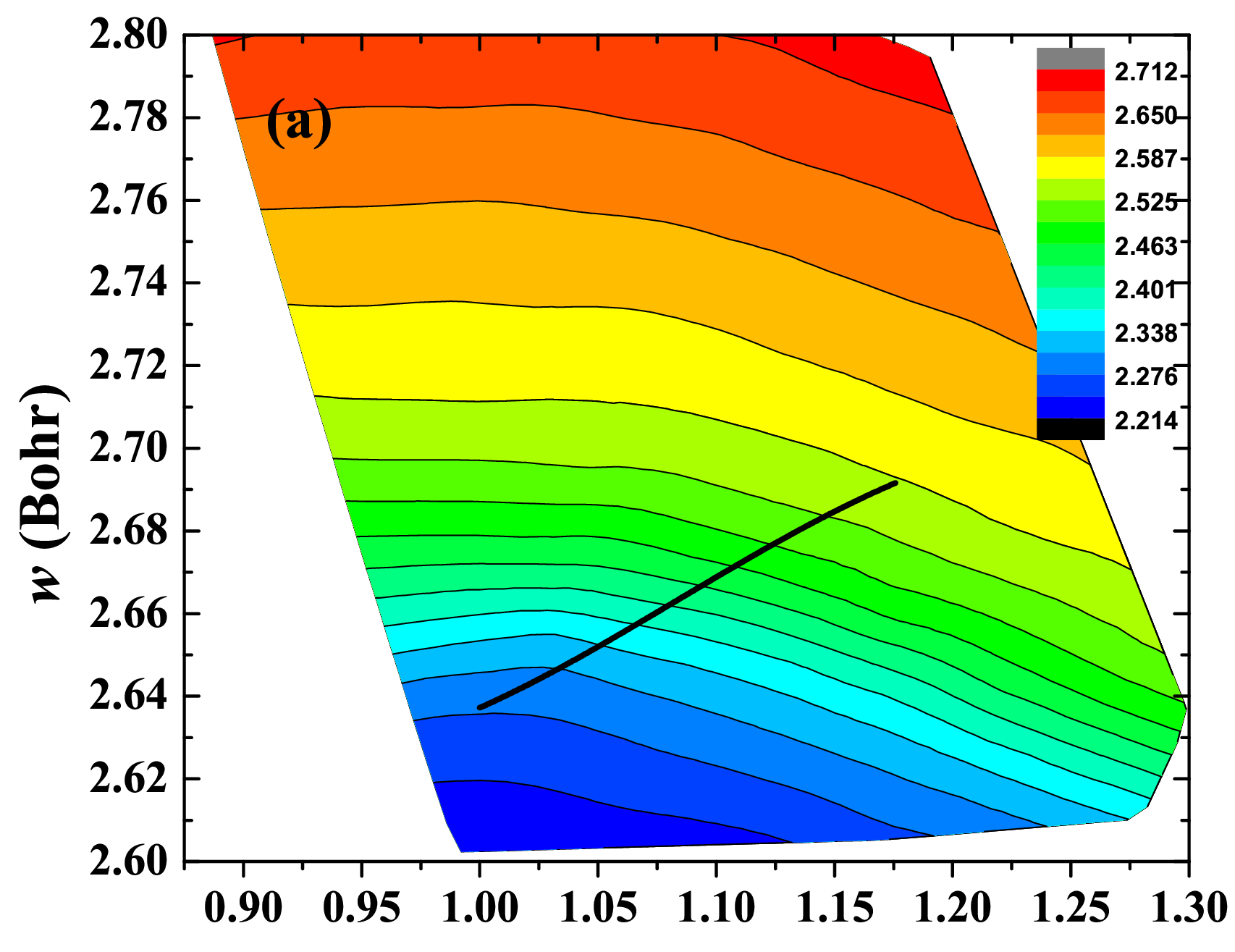}} &
\resizebox{!}{4cm}{\includegraphics[clip]{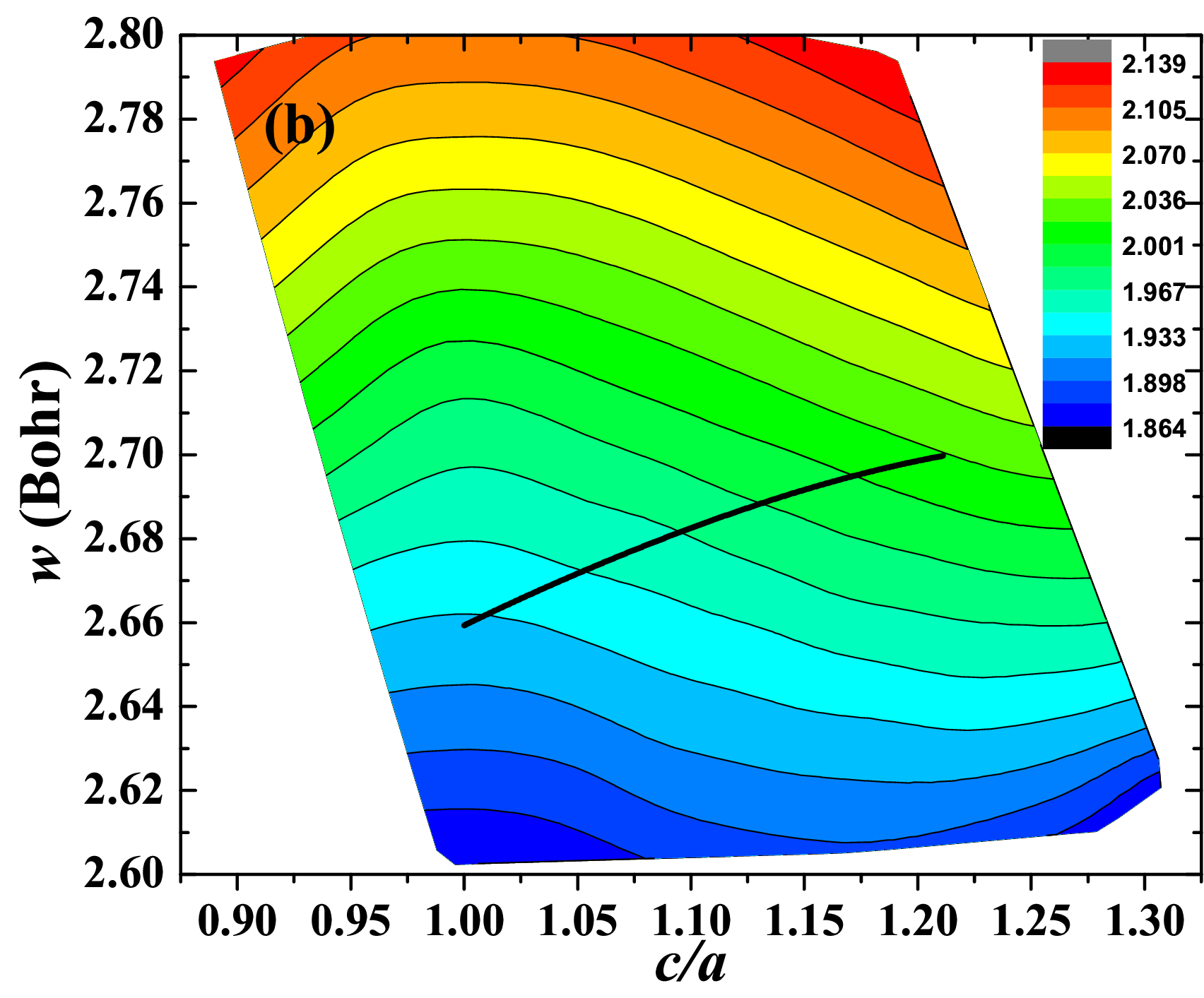}}\\
\resizebox{!}{4cm}{\includegraphics[clip]{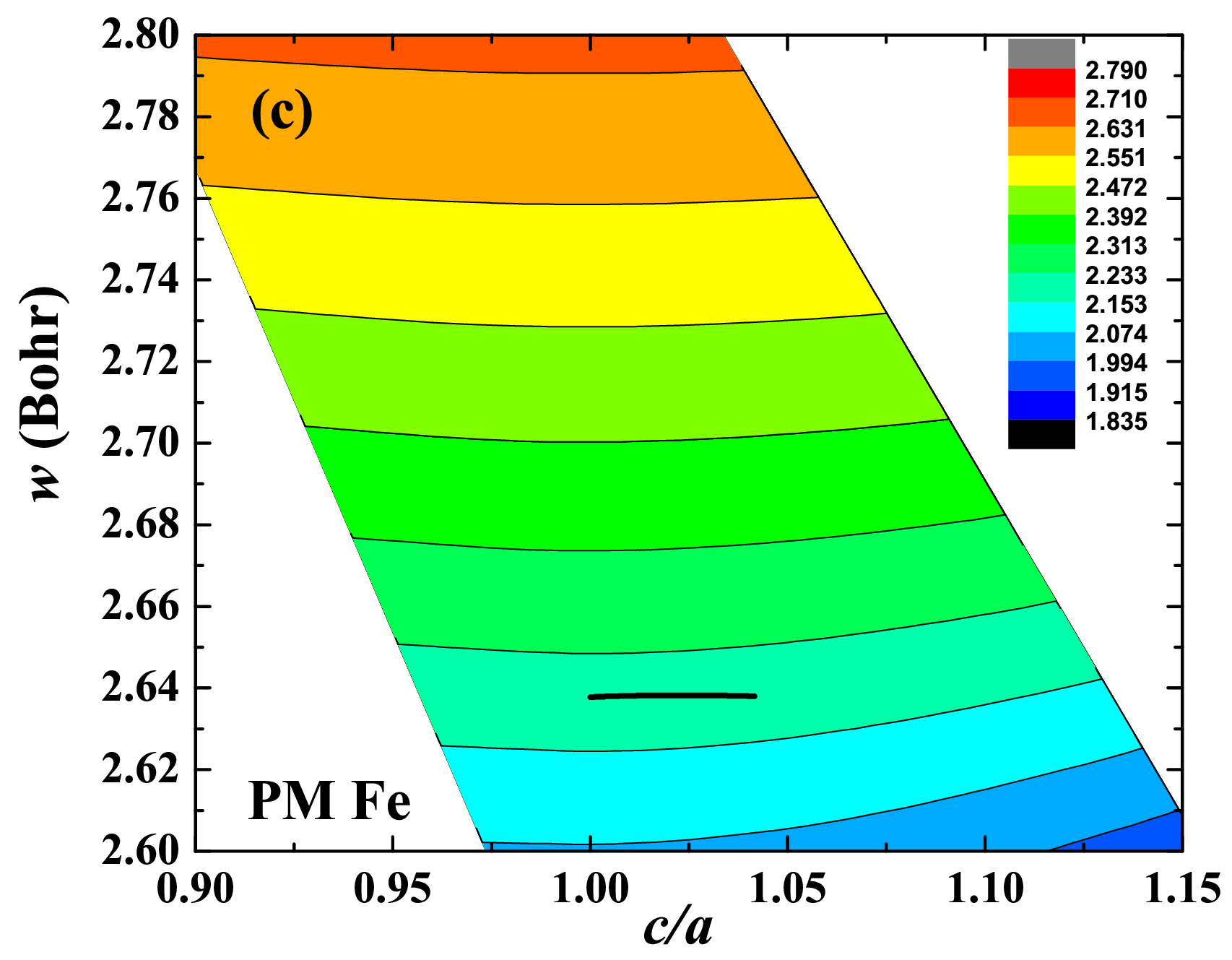}} &
\resizebox{!}{4cm}{\includegraphics[clip]{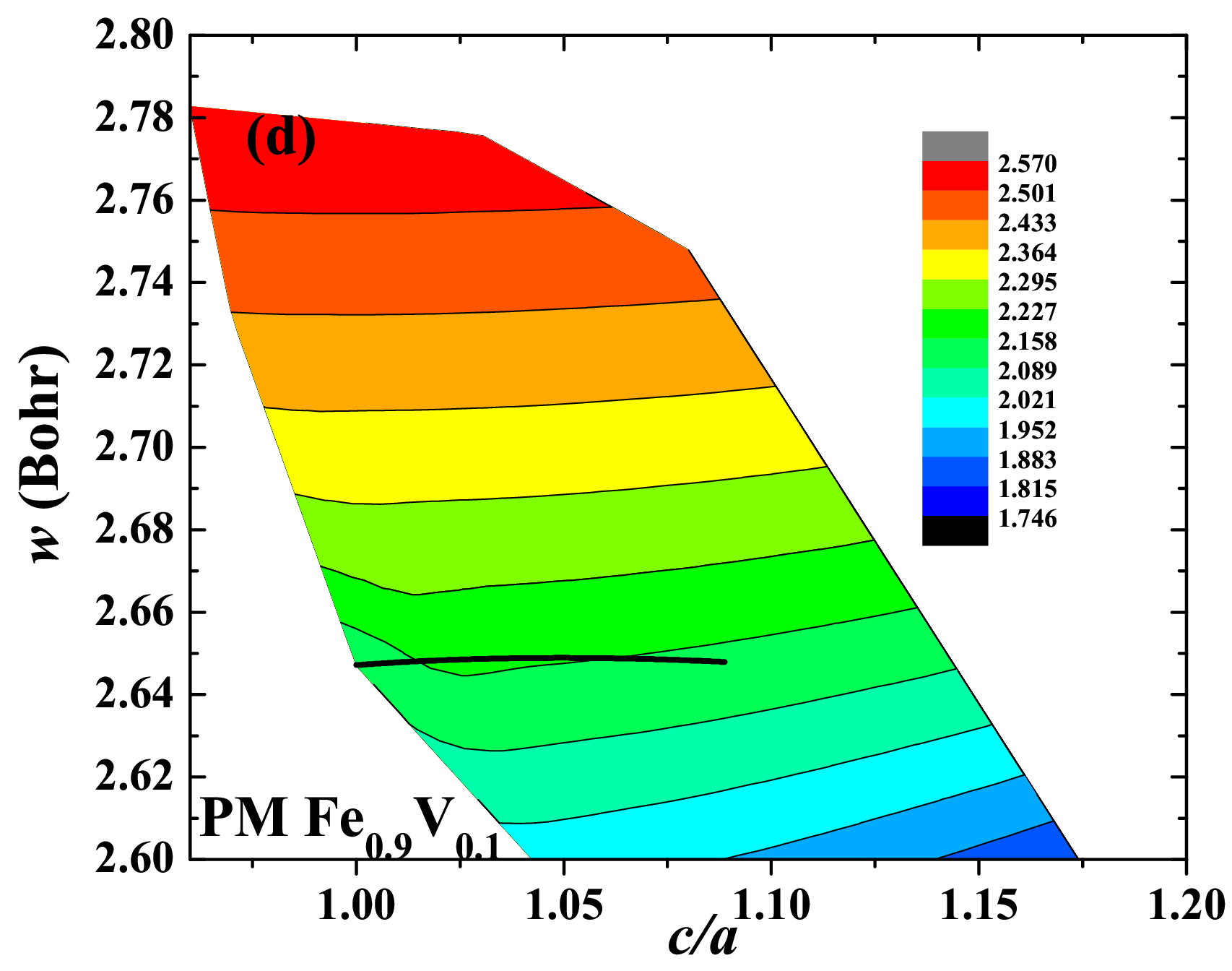}}\\
\end{tabular}
\newline
\caption{\label{fig:5}Magnetic moments (in units of $\mu_{\textrm{B}}$) for pure Fe and Fe$_{0.9}$V$_{0.1}$ random alloy as functions of the tetragonal axial ratio ($c/a$) and Wigner-Seitz radius (\emph{w} in Bohr). Panels (a) and (b) are for the magnetically ordered state and panels (c) and (d) for the disordered state (local magnetic moment).
The black solid lines show the uniaxial deformation paths in the range $\epsilon$ : 0  - $ \epsilon_{\textrm{m}}$.}
\end{center}
\end{figure*}

To achieve a better understanding of the matter,
we chose Fe$_{0.9}$V$_{0.1}$ as a representative of all investigated binary alloys in the following and distinguish 'magnetic moments' in magnetically ordered states from 'local magnetic moments' in the PM state according to common practice.
The magnetic moments and local magnetic moments of Fe and Fe$_{0.9}$V$_{0.1}$ in both magnetic states, respectively, are shown as contour plots in Fig.~\ref{fig:5} as functions of the tetragonal axial ratio ($c/a$) and the Wigner-Seitz radius ($w$).
We also highlighted the corresponding uniaxial deformation path in the range $\epsilon$ : 0 - $\epsilon_{\textrm{m}}$.

Three important insights can be gained on the basis of Fig.~\ref{fig:5}. First, the Wigner-Seitz radius (and thus the atomic volume) has a more pronounced influence on the (local) magnetic moments than the structural change ($c/a$), a feature that can be clearly associated with the contour lines.
Second, the volume along the deformation path from $\epsilon=0$ to $\epsilon_{\textrm{m}}$ hardly changes for PM Fe and Fe$_{0.9}$V$_{0.1}$, implying that the local magnetic moment remains nearly constant during tension. For FM Fe and ferrimagnetic Fe$_{0.9}$V$_{0.1}$ on the other hand, the volumes increase with $\epsilon$ along the indicated uniaxial deformation paths resulting in an increase of the magnetic moments.
Third, regarding the effect of alloying in Figs.~\ref{fig:5}(a) and~(b), the magnetic moment increase ($\Delta \mu$) for FM Fe along the deformation path from $\epsilon=0$ to $\epsilon_{\textrm{m}}$ is much larger than the one obtained for ferrimagnetic Fe$_{0.9}$V$_{0.1}$. Namely, $\Delta \mu$ is $0.3\,\mu_{\textrm{B}}$ for Fe and $0.08\,\mu_{\textrm{B}}$ for Fe$_{0.9}$V$_{0.1}$.

\begin{figure}
 \begin{center}
 \resizebox{\columnwidth}{!}{\includegraphics[clip]{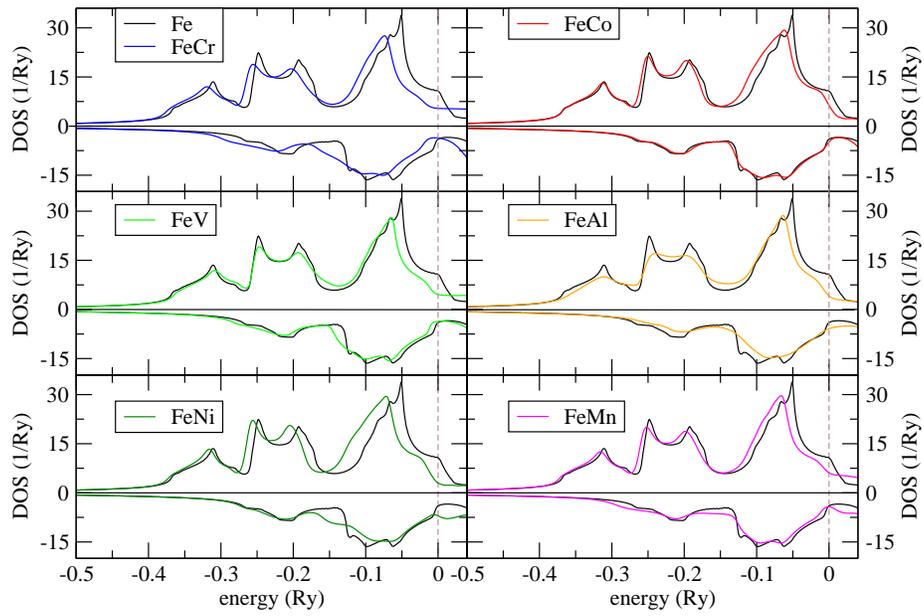}}
\caption{\label{fig:DOS}The total DOS for bcc FM Fe and FFM Fe-based alloys (with 10 at.\,\% solute concentration, 5 at.\,\% in the case of Mn) at their equilibrium volumes. The vertical dashed lines indicate the Fermi level. The DOS of Fe is shown in all panels to ease the comparison. The DOSs of the majority (minority) spin channel have positive (negative) sign.}
\end{center}
\end{figure}

The peculiar difference in the physical behavior during tension, namely that an increase of the magnetic moment is energetically more favorable in pure Fe than in any of the presently investigated Fe binaries, is mainly due to an alloying-induced shift of Fe single-particle states in the majority spin band to lower eigenenergies~\cite{Li:2014}.
This behavior becomes evident from the density of states (DOS) of bcc FM Fe and FFM Fe-based alloys depicted in Fig.~\ref{fig:DOS}: the Fermi level in bcc Fe sits at the bottom of the pseudo gap in the minority band and at the shoulder of the majority band dominated by $t_{2\textrm{g}}$ states. This shoulder 
is shifted to lower energies when adding any of the considered alloying elements as shown in Fig.~\ref{fig:DOS}, while the pseudo gap in the minority spin band is pinned to the Fermi energy.
This alloying-induced effect on the electronic structure of Fe renders the magnetic response of the Fe host during uniaxial tension from the weak towards a stronger FM behavior, and is reflected in the magnetic moment change along the deformation paths as discussed above.

As pointed out by us in Ref.~\cite{Li:2014}, the large increase of the magnetic moment in FM Fe and dilute FFM Fe-based alloys along the uniaxial deformation path leads to the failure of the proposed correlation~\eqref{eq:stress:strdiff} for FFM Fe-based binaries.
In an attempt to disentangle the magnetic effect on the ITS from the chemical effect, we analyzed the excess magnetic pressure of itinerant magnets (proportional to the change of the squared magnetic moment normalized by volume) and demonstrated that the increasing magnetic moment upon lattice distortion induces considerable excess magnetic pressure that partially overrides the pure chemical trend. We also showed that the SED model~\eqref{eq:stress:strdiff} is able to capture the pure chemical trend on the ITS (by suppressing the excess magnetic pressure in the calculations) for all presently considered FFM Fe-based binaries.

Returning finally to PM Fe and the Fe-based alloys, we recall that the local magnetic moment remains largely unperturbed along the uniaxial deformation path. This indicates that, on the one hand, lattice tension induces a negligible excess magnetic pressure and, on the other hand, the alloying trend is mostly governed by the chemical effect. Thus, the compositional effect on the ITS is captured by the SED model as a consequence of a small volume change during uniaxial tension and a negligible excess magnetic pressure.
In the next section, we establish a strong correlation between the ITS and the cubic shear elastic constant $C'$ for the PM Fe alloys that allows corroborating the predominantly chemical nature of the observed alloying trend.

\subsection{\label{sec:correlation}Correlation between ITS and mechanical properties}

In this part, first we investigate a possible correlation between the tetragonal shear elastic constant $C'$ and the present ITS of Fe-based alloys. Then we consider hardenability as a measure of the capability of steels to form martensite.

The shear constant $C'$ is often used to describe the structural stability of a cubic solid under small volume-conserving tetragonal deformation. The ITS, on the other hand, characterizes the mechanical strength of materials at large strain taking into account the Poisson effect (volume change). Due to the distinct strain regimes, a correlation between both mechanical parameters may, thus, not be obvious.
Nevertheless, assuming an analytic, sinusoidal tensile stress-strain curve, Frenkel and Orowan derived a proportionality between the ITS and Young's modulus evaluated in the [001] direction ($E_{[001]}$)~\cite{Frenkel:1926,Orowon:1949}. Since there is a simple relationship between $E_{[001]}$ and $C'$, \emph{viz.}
\begin{equation}
 E_{[001]} = 6C'\frac{B}{C_{11}+C_{12}},
\end{equation}
the SED might indeed correlate with $C'= (C_{11}-C_{12})/2 $, if the compositional effect on $C'$ outweighs the one on the fraction (the correlation is strongest for $2C_{12}\gtrsim C_{11}$). In the previous equation, $B=(C_{11}+2C_{12})/3$ denotes the bulk modulus and $C_{ij}$ are the elastic constants.
Additional motivation is given by the established correlation between $|E_{\textrm{fcc}} - E_{\textrm{bcc}}|$ and $|C'|$ for the transition metal elements family~\cite{Soederlind:1994c}. By virtue of the SED model [Eq.~\eqref{eq:stress:strdiff}] one may then attempt to correlate the ITS with $C'$ if both quantities correlate with $|E_{\textrm{fcc}} - E_{\textrm{bcc}}|$.
It has to be noted, however, that the physical information contained in ground state elastic mechanical parameters like $C'$ discards several phenomena essential to the ideal strength arising from the nature of atomic bonding at large strain, such as bifurcation (branching) of the deformation path and the determination of the failure mode, or volume and lattice relaxations.

\begin{figure}[tbh]
\begin{center}
\begin{tabular}{@{}c@{}}
\resizebox{0.5\columnwidth}{!}{\includegraphics[clip]{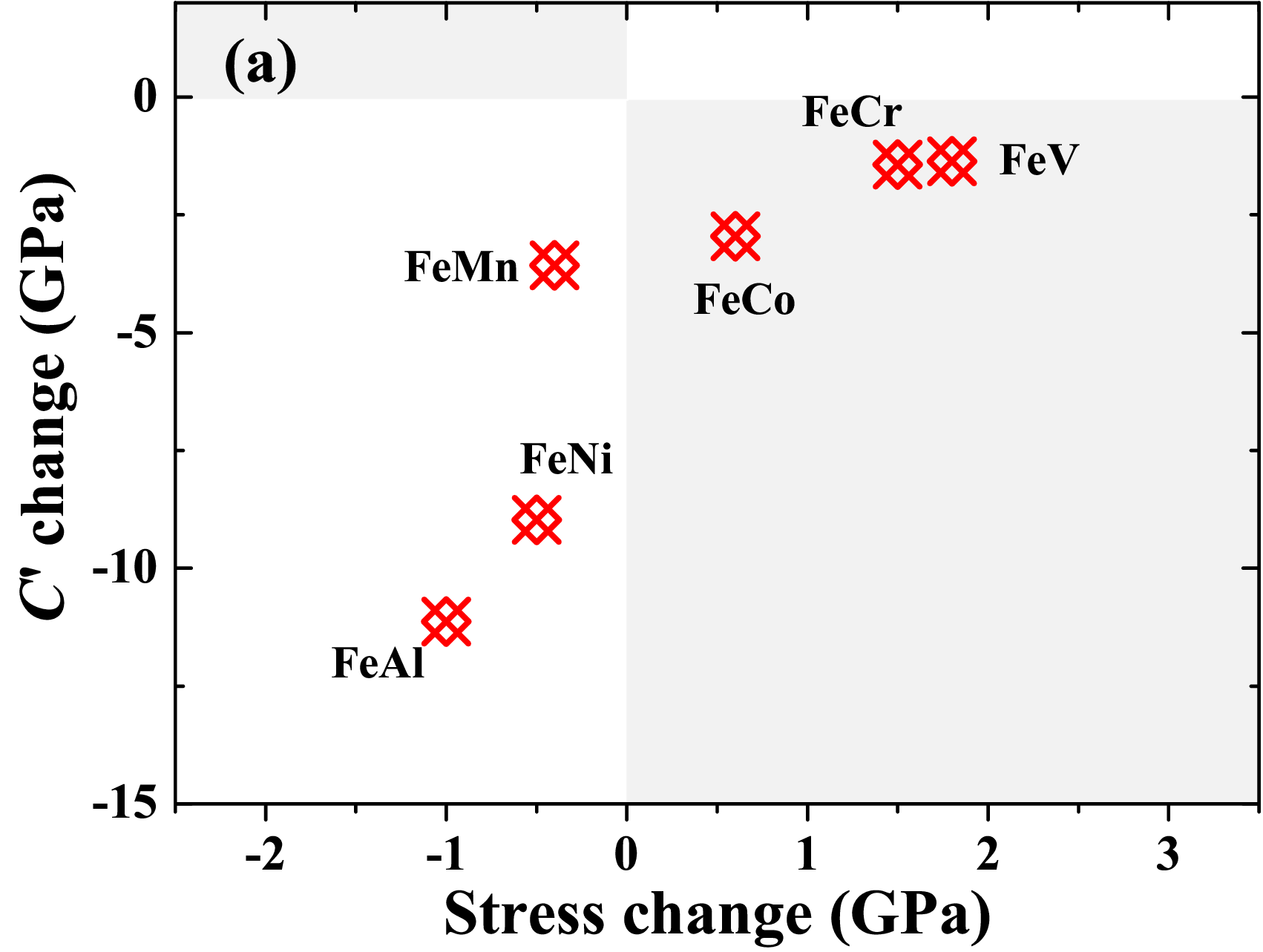}}\\
\resizebox{0.5\columnwidth}{!}{\includegraphics[clip]{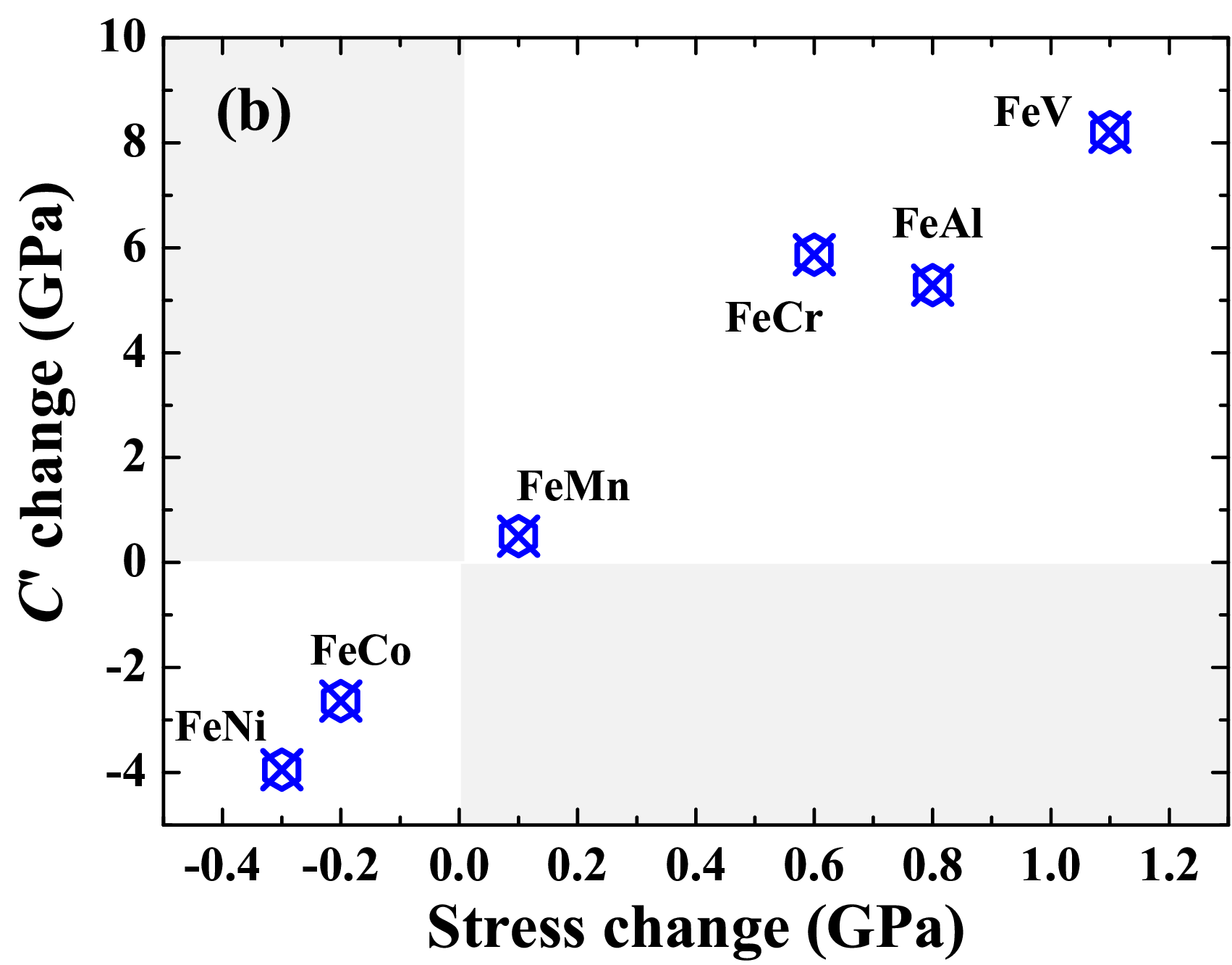}}\\
\end{tabular}
\caption{\label{fig:FM_PM}The change of the tetragonal shear elastic constants $C'$ versus the change of the ITS for the Fe$_{1-x}M_{x}$ alloys in (a) the FFM state for a concentration increase from 0\,\% to 5\,\% and in (b) the PM state for a concentration increase from 0\,\% to 10\,\%, respectively. Data in unshaded areas affirm a correlation.}
\end{center}
\end{figure}

Indeed, the composition dependence of the maximum stress on the primary tetragonal deformation path in ternary V-Cr-Ti alloy was found to correlate with $C'$ and $E_{\textrm{fcc}} - E_{\textrm{bcc}}$ (the ITS occurs on the secondary orthorhombic path)~\cite{Xiaoqing:2013}.
The situation is, however, different and more complex for the presently considered Fe-based alloys: we tested the possible correlation between the alloying effect on $C'$, \emph{i.e.}, $\Delta C' = C'(x) - C'(x_0)$ where values were taken from Refs.~\cite{Zhang:2010,Zhang:2012PM}, and the one on the ITS for FM and PM states and shown the results in Fig.~\ref{fig:FM_PM}.

\begin{figure}[hct]
\begin{center}
\resizebox{\columnwidth}{!}{\includegraphics[clip]{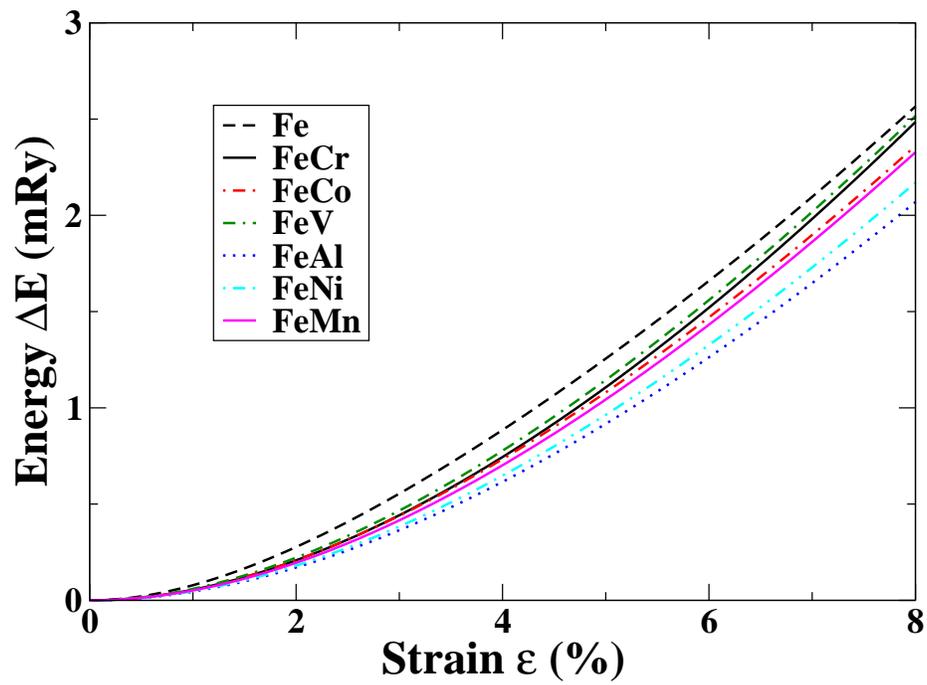}}
\caption{\label{fig:enengystrain}The uniaxial strain energies of FM Fe and FFM Fe$_{1-x}M_{x}$ alloys with 5\,\% solute concentration as a function of strain.}
\end{center}
\end{figure}

From Fig.~\ref{fig:FM_PM}(a), one may conclude that there is no direct correlation between the ITS and $C'$ for the alloying effect of the FFM Fe-based alloys. This finding may be somewhat expected at this point recalling the limited applicability of the SED model as discussed in the previous sections.
Looking at our energy-strain curves in Fig.~\ref{fig:enengystrain} for the intermediate strain region ($\epsilon$ up to approximately 3\,\% - 4\,\%, and $c/a$ up to approximately $1.04$ - $1.05$), the curvature of the Fe curve is, however, larger than those of the Fe-based alloys. In other words, alloying produces the same effect on $C'$ (results from Refs.~\cite{Zhang:2010,Zhang:2012PM}) and on the deformation behavior at intermediate strains.
For strains larger than $\approx$ 3\,\% - 4\,\%, the slopes of the energy-strain curves follow the trend of the ITSs (Fig.~\ref{fig:alloysstress}).

By contrast, Fig.~\ref{fig:FM_PM}(b) reveals that the ITS change correlates strongly with the $C'$ change for the PM Fe-based alloys.
As an application of the established correlation~\ref{fig:FM_PM}(b), we may substantiate the predominantly chemical origin of the alloying trend on the ITS of PM Fe. By means of fixed moment calculations for the local magnetic moment of PM Fe, Dong \emph{et al.} showed that the magnitude of $C'$ for bcc PM Fe decreases considerably if the local moment drops below its equilibrium value~\cite{Dong:2015}. Our investigations reveal that 10\,\% Co and Ni increase the local magnetic moment of the PM Fe host (Co itself is polarized, while Ni is not) whilst all other solutes reduce the local magnetic moment of Fe.
If there were a strong magnetic origin of the alloying effect on the ITS (a governing coupling between the magnitude of the ITS and the local magnetic moment of the PM Fe host), one might expect by virtue of correlation~\ref{fig:FM_PM}(b) and Dong \emph{et al.}'s results that the ITS of the PM Fe-based alloys increased for Co and Ni and decreased for the other solutes.
This is, however, in contraction to the present findings (Fig.~\ref{fig:alloysstress}) indicating the predominantly chemical origin of the alloying trend on the ITS.

The ability to harden a steel not only at the surface but also in considerable depth in the interior of the specimen by quenching is reflected by the hardenability of a steel.  The hardness of quenched steel depends on the microstructure, \emph{e.g.}, the formation of the martensite phase. A larger hardenability of a steel indicates a larger ability to harden the steel by martensite formation.
In martensite, interstitially trapped C (carbon) results locally in a deformation of the bcc lattice to the bct lattice with $c/a =1+(0.046 \pm 0.001)n$, where $n$ stands for the amount of C in expressed in weight percent~\cite{Kurdjumov:1976}. If this deformation is softer, it is easier to form martensite by quenching. According to our data (Fig.~\ref{fig:enengystrain}), all alloying elements soften the bcc to bct lattice distortion since their uniaxial strain energies are lower than the one of FM Fe for intermediate strains (up to approximately $\epsilon=8\,\%$, corresponding $c/a\approx 1.1$; note that for $\epsilon>8\,\%$, the strain energy of Fe is not the largest anymore). This suggests that martensite formation may be favored if FM Fe is alloyed with Al, V, Cr, Mn, Co, or Ni, and hence these solutes are expected to increase the hardenability of steel. The largest effect is obtained for Al and Ni followed by Mn, Co, Cr, and V.
Available experiments indicate that the hardenability of steels is greatly influenced by chemical composition. In particular, Cr, Ni, Al, Mn, and V increase the hardenability of low carbon steels~\cite{Siebert:1977}, which is in line with our prediction.

\section{Conclusions}
The ideal tensile strengths of bcc ferromagnetic and paramagnetic Fe and ferrite-rich Fe$_{1-x}M_{x}$ (\emph{M}=Al, V, Cr, Mn, Co, or Ni) random alloys in the $[001]$ direction were investigated using the all-electron exact muffin-tin orbitals method in combination with the coherent-potential approximation. The present ideal tensile strength value of ferromagnetic Fe and the branching point from the primary tetragonal deformation path to the secondary orthorhombic deformation path agree well with full-potential results and previously published calculations, which confirms that our methodology has the accuracy needed for such simulations.
For the Fe-based alloys in the ferro-/ferrimagnetic state, we found that the ideal tensile strength increases with increasing amount of Cr, Co, or V and decreases with Ni or Al addition to pure Fe. Manganese shows a small but non-monotonic alloying behavior.
Compared to ferromagnetic Fe, paramagnetic Fe possesses a significantly lower ideal tensile strength. The relative magnitude of the ITS per solute concentration can be altered more strongly in the paramagnetic state, for which we found that Al and the considered transition metal elements with $d$-occupation number smaller than Fe increase the ITS of paramagnetic Fe, while Co and Ni possessing larger $d$-band fillings lower it.
All binary alloys considered here are intrinsically brittle and fail by cleavage of the $(001)$ planes under uniaxial tensile loading in the $[001]$ direction in both magnetic phases.

Unlike the nonmagnetic bcc V-based alloys~\cite{Xiaoqing:2013} and the paramagnetic bcc Fe-alloys, the employed ideal tensile strength model based on constant volume structural energy differences can not capture the alloying effect on the ideal tensile strength for the ferro-/ferrimagnetic bcc Fe-alloys.
We identified the interplay between magnetism and volume change in response to uniaxial tension as the main driving force for the distinct behavior of Fe. Making use of the established concept of magnetic pressure, we showed that paramagnetic Fe alloys are characterized by a small volume change during uniaxial tension and a negligible excess magnetic pressure. This indicates that the compositional effect on the ITS is mostly governed by the chemical effect of the solute. By contrast, ferro-/ferrimagnetic Fe-alloys exhibit an increasing magnetic moment upon lattice distortion inducing a considerable excess magnetic pressure that partially overrides the chemical trend.

We established a strong, direct correlation between the compositional effect on the ideal tensile strength and the one on the shear elastic constant $C'$ for the investigated paramagnetic bcc Fe alloys. A similar correlation does not hold when Al, V, Cr, Mn, Co, or Ni are alloyed to ferromagnetic Fe. We discussed that all solutes are expected to increase the hardenability of ferritic steel.

The present results offer a consistent starting point for further theoretical modeling of the micro-mechanical properties of Fe-based alloys in various magnetic states. Based on these achievements, we conclude that  modern computational alloy theory provides an efficient and accurate theoretical tool to design the mechanical strength of ferro-/ferrimagnetic bcc random solid solutions.

\section{ACKNOWLEDGEMENTS}
The Swedish Research Council, the Swedish Steel Producers' Association, the European Research Council, the China Scholarship Council and the Hungarian Scientific Research Fund (research project OTKA 84078 and 109570), and the National Magnetic Confinement Fusion Program of China (2011GB108007) are acknowledged for financial support.
S.\! S.\!  gratefully acknowledges the Olle Erikssons stiftelse f\"or materialteknik.
The simulations were performed on resources provided by the Swedish National Infrastructure for Computing (SNIC) at the National Supercomputer Centre in Link\"oping.

\end{document}